\documentclass[aps,twocolumn,preprintnumbers,amsmath,amssymb,superscriptaddress]{revtex4-1}

\usepackage{graphicx}
\usepackage{dcolumn}
\usepackage{epsfig}
\usepackage{color}
\usepackage{ulem}

\begin{document}

\title{Coupling of Higgs and Leggett modes in nonequilibrium superconductors}

\author{H. Krull}
\email{holger.krull@tu-dortmund.de}
\affiliation{Lehrstuhl f\"{u}r Theoretische Physik I, Technische Univerit\"at Dortmund, 
Otto-Hahn Stra\ss{}e 4, 44221 Dortmund, Germany}
\author{N. Bittner}
\email{n.bittner@fkf.mpg.de}
\affiliation{Max-Planck-Institut f\"{u}r Festk\"{o}rperforschung,
Heisenbergstra\ss{}e 1, D-70569 Stuttgart, Germany}
\author{G. S. Uhrig}
\email{goetz.uhrig@tu-dortmund.de}
\affiliation{Lehrstuhl f\"{u}r Theoretische Physik I, Technische Univerit\"at Dortmund, 
Otto-Hahn Stra\ss{}e 4, 44221 Dortmund, Germany}
\author{D. Manske}
\email{d.manske@fkf.mpg.de}
\affiliation{Max-Planck-Institut f\"{u}r Festk\"{o}rperforschung,
Heisenbergstra\ss{}e 1, D-70569 Stuttgart, Germany}
\author{A. P. Schnyder}
\email{a.schnyder@fkf.mpg.de}
\affiliation{Max-Planck-Institut f\"{u}r Festk\"{o}rperforschung,
Heisenbergstra\ss{}e 1, D-70569 Stuttgart, Germany}

\date{\today}
\maketitle

\noindent
{\bf 
Collective excitation modes
are a characteristic feature of symmetry-broken phases of matter.
For example, superconductors exhibit an amplitude Higgs mode and
a phase mode, which are the radial and angular excitations
in the Mexican-hat potential of the free energy.
In two-band superconductors there exists in addition a Leggett phase mode, which corresponds to 
collective fluctuations of the interband phase difference.
In equilibrium systems  amplitude and phase modes are decoupled, since they are mutually orthogonal excitations.
The direct detection of  Higgs and Leggett modes by linear-response measurements is  challenging,
because they are often overdamped
and do not couple directly to the electromagnetic field.
In this work, using numerical exact simulations  we show for the case of  two-gap superconductors, that
optical pump-probe experiments excite
both  Higgs and Leggett modes out of equilibrium.
We find that this non-adiabatic excitation process introduces a  strong interaction between the collective modes.
Moreover, we predict that the coupled Higgs and Leggett modes  are clearly visible in 
the pump-probe absorption spectra as oscillations at their respective frequencies.  }

 %%%%%%%%%%%%%%%%%%%
\begin{figure}[t!]
\includegraphics[width=0.75\columnwidth]{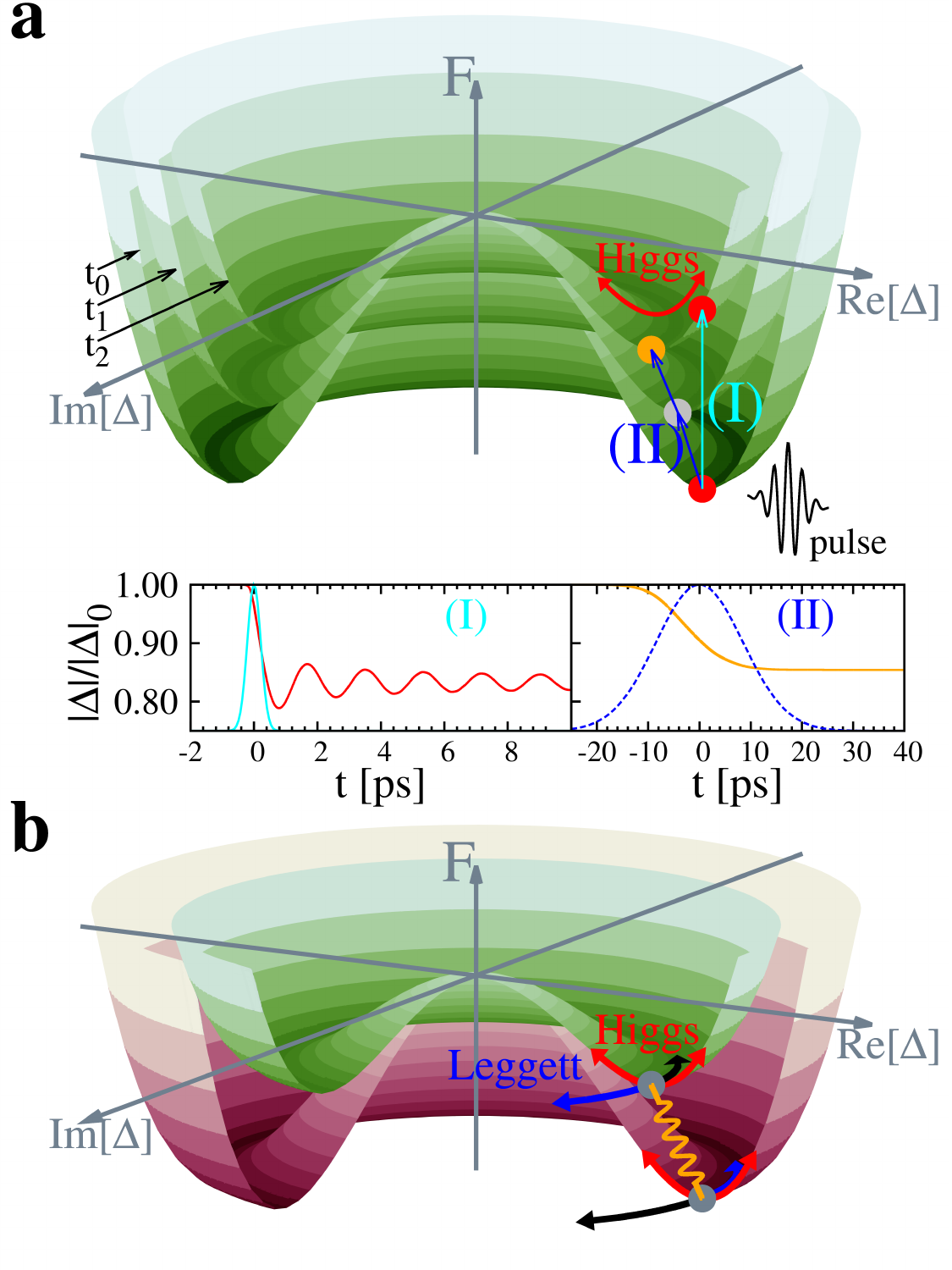}  
\caption{ \label{mFig0}
\textbf{Illustration of Leggett and Higgs modes.} 
(\textbf{a})~Illustration of the excitation process for a one band superconductor. 
The pump laser pulse modifies the free energy $\mathcal{F}$ on different time scales depending on the pulse duration~$\tau$. 
For $\tau \gg h / (2 | \Delta |)$ the superconductor can follow the change in $\mathcal{F}$ adiabtatically, 
resulting in a monotonic lowering of the order parameter $| \Delta |$ [inset (II)]. For short pulses with $\tau  \lesssim h / (2 | \Delta |)$,
on the other hand,
the superconductor is excited in a non-adiabatic fashion, which results in oscillations of $| \Delta |$ about the new minimum of $\mathcal{F}$ [inset (I)]. 
The blue and cyan lines in the two insets represent the Gaussian profiles of the pump pulses.
(\textbf{b})~Effective free-energy landscape $\mathcal{F}$
for a two-gap superconductor, with green and red representing 
 the Mexican-hat potentials of the smaller and larger gaps, respectively.  The amplitude Higgs  modes and
 the phase modes are indicated by red and blue/black arrows, respectively.
The Leggett mode corresponds to
out-of-phase  fluctuations of the phase difference between the two gaps.
}
\end{figure}
%%%%%%%%%%%%%%%%%%%

%%%%%%%%%%%%%%%%%%%
\begin{figure*}[t!]
\includegraphics[width=0.95\textwidth]{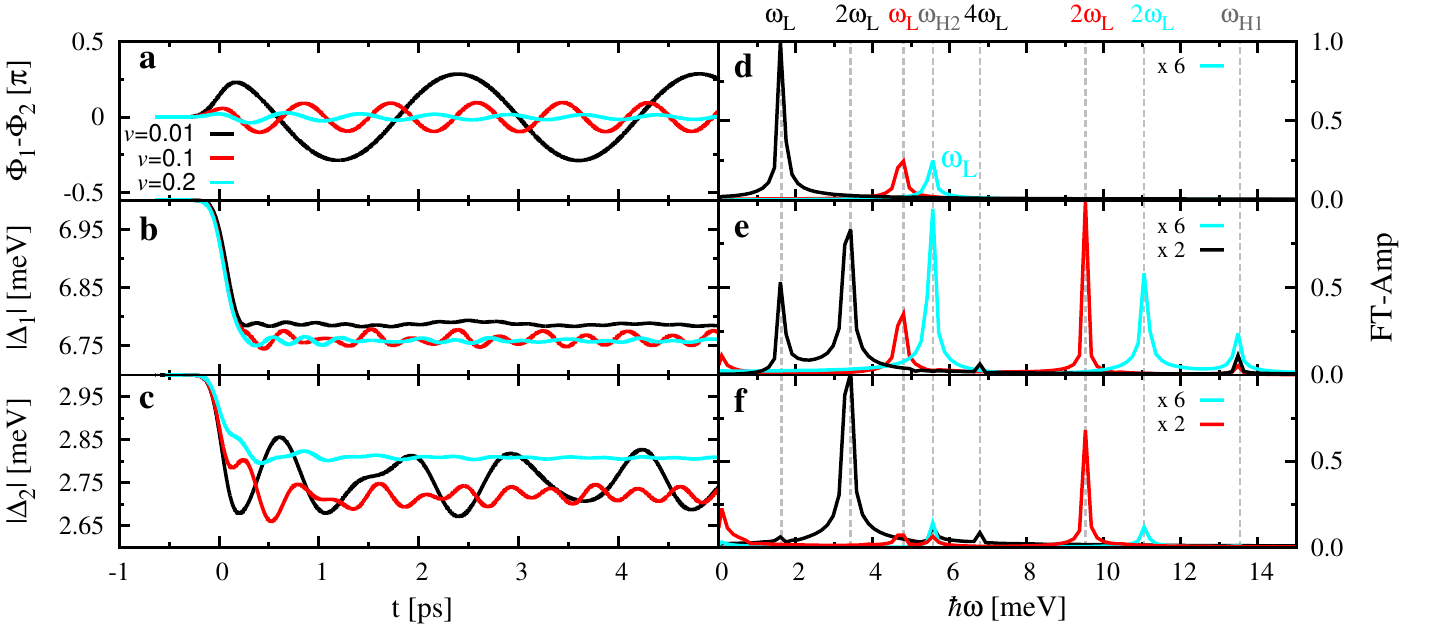}  
\caption{\label{mFig1}
\textbf{Leggett phase mode and amplitude Higgs mode oscillations.} 
Numerical simulation of the gap dynamics of a two-gap superconductor after a non-adiabatic excitation by
a short intense laser pulse of width $\tau=0.4$~ps,
pump energy $\hbar \omega_{0}=8$~meV,
and light-field amplitude $|\textbf{A}_0|=10 \cdot 10^{-8}$ Js/(Cm).
(\textbf{a}) Phase difference $\Phi_1 - \Phi_2$ between the two gaps as a function of time $t$ for various interband coupling strengths $v$.
(\textbf{d}) Fourier spectrum of the oscillations in panel (\textbf{a}).
The frequency of the nonequilibrium Leggett mode oscillation is indicated by $\omega_{\textrm{L}}$.
 (\textbf{b}),  (\textbf{c}) Gap amplitudes $| \Delta_1|$ and $| \Delta_2|$ as a function of time $t$ for different interband couplings $v$.
 (\textbf{e}),  (\textbf{f}) Fourier spectra of the amplitude mode oscillations in panels  (\textbf{b}),  (\textbf{c}), which
 display the following frequencies:  $\omega_{\text{H}1}$ and $\omega_{\text{H}2}$ the frequencies of the nonequilibrium Higgs modes 
 of gap $\Delta_1$ and $\Delta_2$, respectively; $\omega_{\textrm{L}}$ the frequency of the nonequilibrium Leggett mode;
 and  higher harmonics of the nonequilibrium Leggett mode denoted by $2 \omega_{\textrm{L}}$ and $4 \omega_{\textrm{L}}$.
}
\end{figure*}
%%%%%%%%%%%%%%%%%%%

Ultrafast pump-probe 
measurements  
have become a key tool 
to probe the temporal dynamics and relaxation  of quantum materials.
This technique has proven to be particularly valuable for
the study of 
order parameter dynamics
in symmetry-broken states, 
such as superconductors~\cite{matsunagaPRL12,matsunagaPRL13,Matsunagascience14,mansartPNAS13,Pashkinscience14,DalScience2013,Beck2013,leitensdorferPRL10,beckDemsarPRL11}, charge-density-waves~\cite{hellmannNatCommun12,gedik_cuprate_CDW_nat_mat_13}, and antiferromagnets~\cite{kimelAFM_nature_04}.
In these experiments the pump  laser pulse excites a high density of quasiparticles above the gap of 
the order parameter, thereby modifying the Mexican-hat potential of the 
free energy  $\mathcal{F}$.  
As a result, the amplitude of the order parameter decreases, reducing the 
minimum of the free energy.
If the pump-pulse induced changes in $\mathcal{F}$ occur on a faster time scale than the intrinsic response time of the symmetry-broken state,
the collective modes  start to oscillate at their characteristic frequencies about the new free-energy minimum  [see Fig.~\ref{mFig0}(a)].
This non-adiabatic excitation mechanism has recently been demonstrated for the  amplitude  
Higgs mode of the one-gap superconductor NbN.
It has been shown, both theoretically~\cite{amin2004,volkov1974,Schny11,Papen07,Papen08,Yuz06,Yuz05,YuzJofPhys05,zachmannNJP13,Krull14,Yuz14,Krullphd,kemperArxiv14,peronaciArxiv15}  and experimentally~\cite{matsunagaPRL12,matsunagaPRL13,Matsunagascience14}, that a short intense laser pulse
of length $\tau$ much shorter than  the dynamical time scale of the superconductor $\tau_{\Delta} \simeq h / (2 | \Delta |)$
induces oscillations in the order parameter amplitude at the frequency $\omega_{\text{H}} =  2  \Delta_{\infty}  / \hbar$, 
with $\Delta_{\infty}$ the asymptotic gap value.

While nonequilibrium collective modes in conventional single-gap superconductors are well understood,
the investigation of collective excitations in unconventional nonequilibrium superconductors with multiple gaps,
such as MgB$_2$ or  iron pnictides,
is still in its infancy~\cite{Akba13,dzeroPRB15,Demsar2003}.
These multicomponent superconductors have a particularly rich spectrum 
of collective excitations~\cite{Leggett66,burnellPRB10,AnishchankaEfetov07}. 
In this paper, we simulate the pump-probe process in a two-gap 
superconductor using a semi-numerical approach based on the 
density-matrix formalism.
This method is exact for mean-field Hamiltonians~\cite{volkov1974,YuzJofPhys05}, captures the coupling between the superconductor and the electromagnetic field
of the pump laser at a microscopic level, and
allows for
the calculation of the pump-probe conductivity, as measured in recent experiments~\cite{matsunagaPRL12,matsunagaPRL13}.
Two-gap superconductors
exhibit besides the amplitude Higgs~\cite{pekkerVarmaReview}  
and the  phase modes~\cite{andersonPR58,bogoliubovBook},
also a Leggett  mode~\cite{Leggett66}, which results from
 fluctuations of the relative phase of the two coupled gaps,
i.e., equal but opposite phase shifts of the two order parameters, see Fig.~\ref{mFig0}(b).
In equilibrium superconductors, the Higgs and Leggett modes are decoupled, since they 
correspond to mutually orthogonal fluctuations. 
In contrast to the phase mode, both Higgs and Leggett modes are charge neutral and therefore 
do not couple directly to the electromagnetic field~\cite{podolskiPRB11}. This has made it difficult to directly detect these 
excitations with standard linear-response type measurements~\cite{Sooryakumar_klein_PRL80,LittlewoodVarma_PRL_81,blumbergPRL07,sacutoPRB14}.

Here, we show that in a pump-probe experiment both   Leggett and   Higgs modes
can be excited out of equilibrium, and directly observed
as oscillations in the absorption spectra at their respective frequencies.
We find that the non-adiabatic excitation process of the pump pulse induces
an intricate coupling between the two charge-neutral modes,
which pushes the frequency of the Leggett mode below the continuum of  two-particle excitations. Moreover,
the  frequencies of the Leggett and Higgs modes and the coupling between them can be 
controlled by the fluence of the pump pulse. Hence, by adjusting the 
laser intensity the  two modes can be brought into resonance, which greatly enhances 
their oscillatory signal in the pump-probe absorption spectra.

\vspace{0.5cm}
 
\noindent
\textbf{\large Results}\\
In this work we use numerical exact simulations in order to study the nonequilibrium response of two-gap superconductors
perturbed by a short and intense pump  pulse.
The Hamiltonian describing the superconductor coupled
to the pump laser field is given by $H = H_{\textrm{BCS}} + H_{\textrm{laser}}$, 
 with the two-band BCS mean-field Hamiltonian
\begin{eqnarray} \label{eq_ham}
H_{\textrm{BCS}} 
= 
H_0
+
\sum_{{\bf k} \in \mathcal{W} } \sum_{l=1}^2
\left( \Delta_l c^{\dag}_{{\bf k}l \uparrow} c^{\dag}_{- {\bf k} l \downarrow}
+
\Delta^{\ast}_l c^{\ }_{- {\bf k} l \downarrow} c^{\ }_{{\bf k}l \uparrow } \right) , \quad
\end{eqnarray}
where  $H_0 = \sum_{{\bf k} l \sigma} \varepsilon_{{\bf k} l} c^{\dag}_{{\bf k} l \sigma } c^{\ }_{{\bf k} l \sigma }$ denotes the normal state Hamiltonian and
$c^{\dag}_{{\bf k} l \sigma}$ creates electrons with momentum ${\bf k}$, band index $l$,
and spin $\sigma$.  
The first sum  in Eq.~\eqref{eq_ham} is
taken over the set $\mathcal{W}$ of momentum vectors with 
$|\varepsilon_{{\bf k} l}| \leq \hbar \omega_c = 50$~meV,  $\omega_c$ being the cut-off frequency.
The gaps $\Delta_1$ and $\Delta_2$ in the two bands are determined at each 
temporal integration step from the BCS gap equations with the 
attractive intraband pairing interactions $V_{1}$ and $V_{2}$ and
the interband coupling $V_{12} = v V_{1}$. 
We fix $V_1$ and $V_2$ such that the gaps
in the initial state
take on the values $\Delta_1 (t_i) = 7$~meV and  $\Delta_2 (t_i) = 3$~meV,
and study the dynamics of the two-gap superconductor as a function of the relative interband coupling $v$. 
$H_{\textrm{laser}}$   represents the
interaction  of the pump laser with the superconductor
 and contains terms linear and quadratic in the vector potential of the laser field,
 which is of Gaussian shape with central frequency $\hbar \omega_{0} = 8$~meV, pulse width $\tau =0.4$~ps, 
 and light-field amplitude $| \textbf{A}_{0}|$.
  We determine the dynamics of Hamiltonian~\eqref{eq_ham} by means of the density matrix approach 
 and solve the resulting  equations of motion using Runge-Kutta integration  (see Methods).

%%%%%%%%%%%%%%%%%%%
\begin{figure}[t!]
\includegraphics[width=0.45\textwidth]{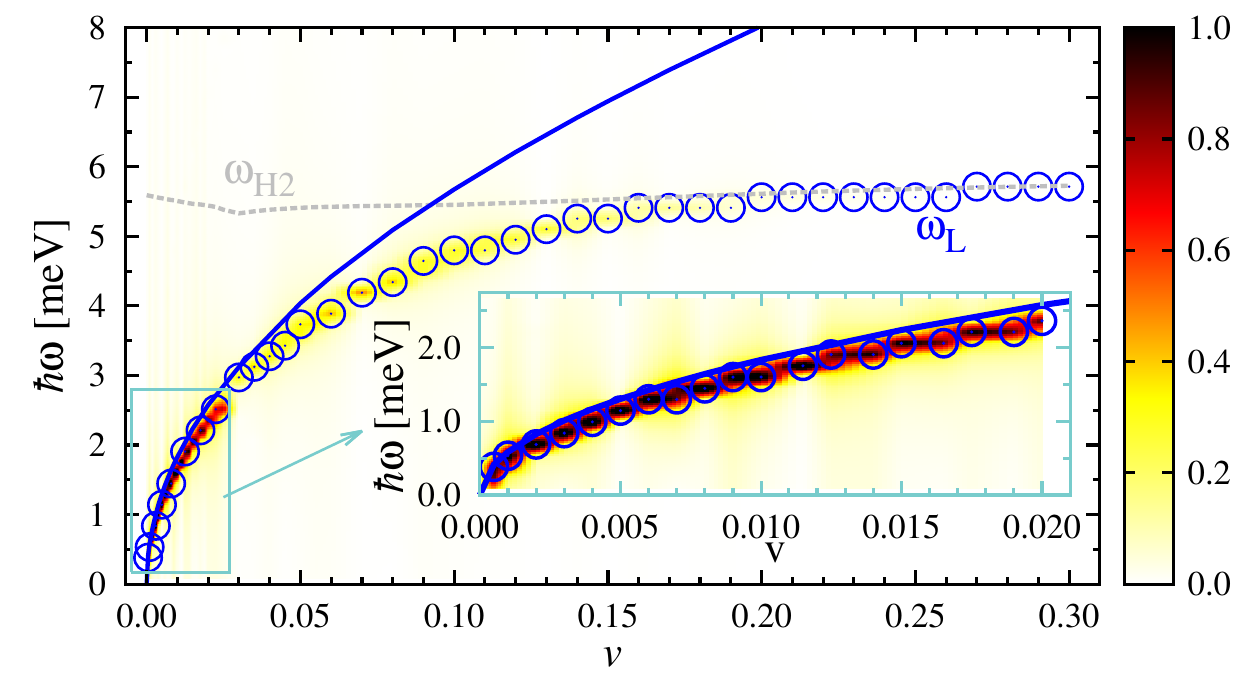}  
\caption{ \label{mFig3}
\textbf{Leggett phase mode oscillations vs.\  relative interband coupling.} 
Fourier spectrum of the phase mode $\Phi_1 - \Phi_2$
 as a function
of relative interband coupling $v$ for a two-band superconductor perturbed by 
 the same laser pulse as in Fig.~\ref{mFig1}. 
The amplitude of the phase fluctuations is indicated by the color scale with 
dark red and light yellow representing the highest and lowest amplitudes, respectively.
The  blue open circles mark the frequency of the nonequilibrium Leggett mode $\omega_{\textrm{L}}$.
The blue solid line represents the frequency of
the equilibrium Leggett mode described by equation~\eqref{equilibriumLeggett}.
The dashed gray line indicates the frequency of the Higgs mode $\omega_{\textrm{H2}}$, which
coincides with 
the boundary to the continuum of
Bogoliubov quasiparticle excitations,   given by twice the asymptotic gap value of the second band $2 \Delta_2^{\infty}$.
The inset shows a zoom-in of the blue frame in the main panel.
}
\end{figure}
%%%%%%%%%%%%%%%%%%%

%%%%%%%%%%%%%%%%%%%
\begin{figure}[t!]
\includegraphics[width=0.45\textwidth]{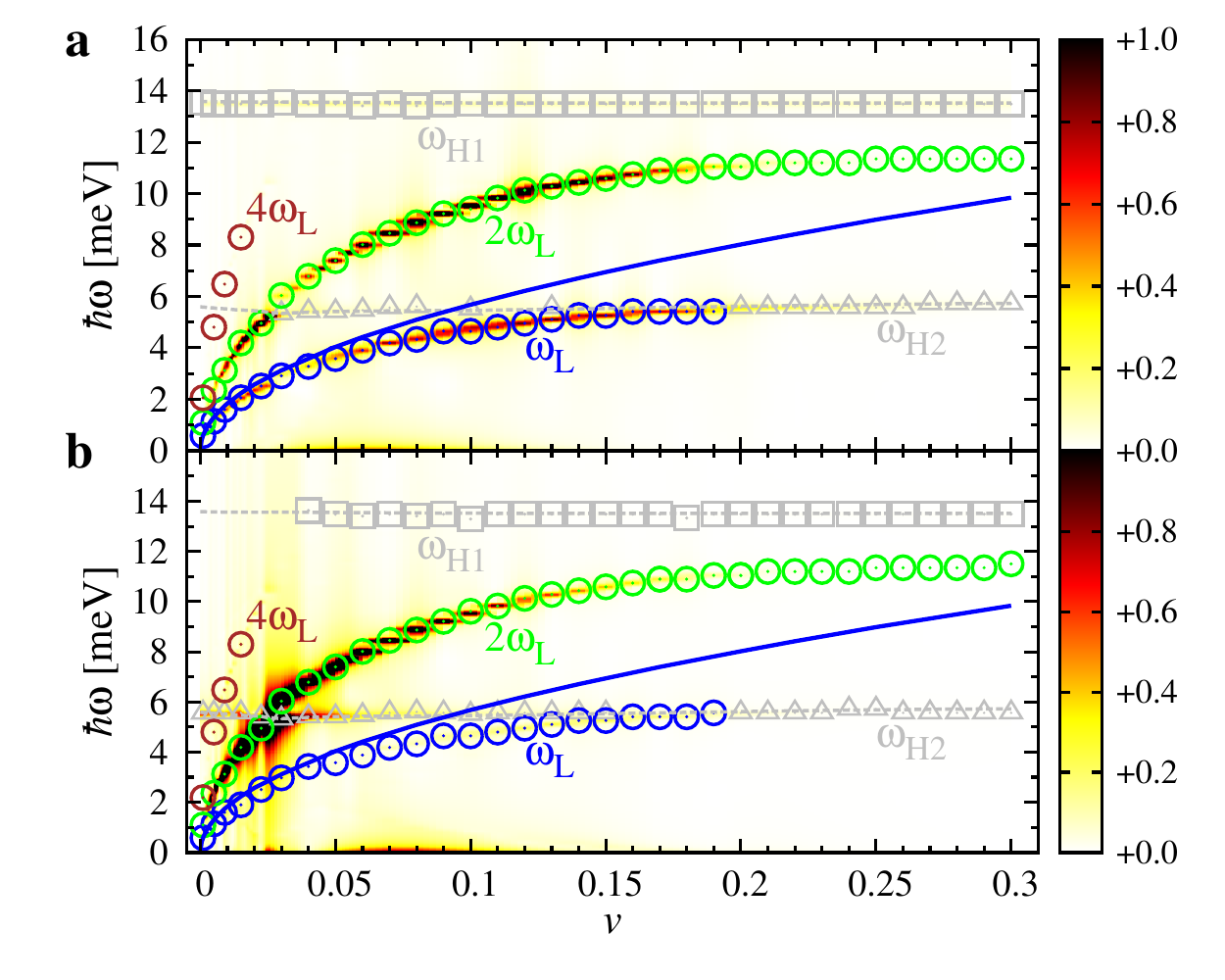} 
\caption{ \label{mFig4}
\textbf{Amplitude mode oscillations vs.\ relative interband coupling.}
Fourier spectrum of the amplitude mode oscillations
as a function of relative interband coupling $v$ for (\textbf{a}) the superconducting gap $\Delta_1$ in
the first band and (\textbf{b}) the superconducting gap $\Delta_2$ in the second band.
The parameters of the laser pump pulse are the same as in Fig.~\ref{mFig1}.
The amplitude of the oscillations is indicated by the color scale with dark red and light yellow
representing the highest and lowest amplitudes, respectively.
The open circles represent the frequencies of the nonequilibrium Leggett mode $\omega_{\textrm{L}}$ and its higher  
harmonics denoted by $2\omega_{\textrm{L}}$ and $4\omega_{\textrm{L}}$.
The frequencies of the nonequilibrium Higgs mode of the first and second band,
$\omega_{\text{H}1}$ and $\omega_{\text{H}2}$,
are indicated by the grey open squares and triangles, respectively.
The blue solid line is the frequency of the equilibrium Leggett mode given by
equation~\eqref{equilibriumLeggett}.}
\end{figure}
%%%%%%%%%%%%%%%%%%%

\begin{figure*}[t!]
\includegraphics[width=0.8\textwidth]{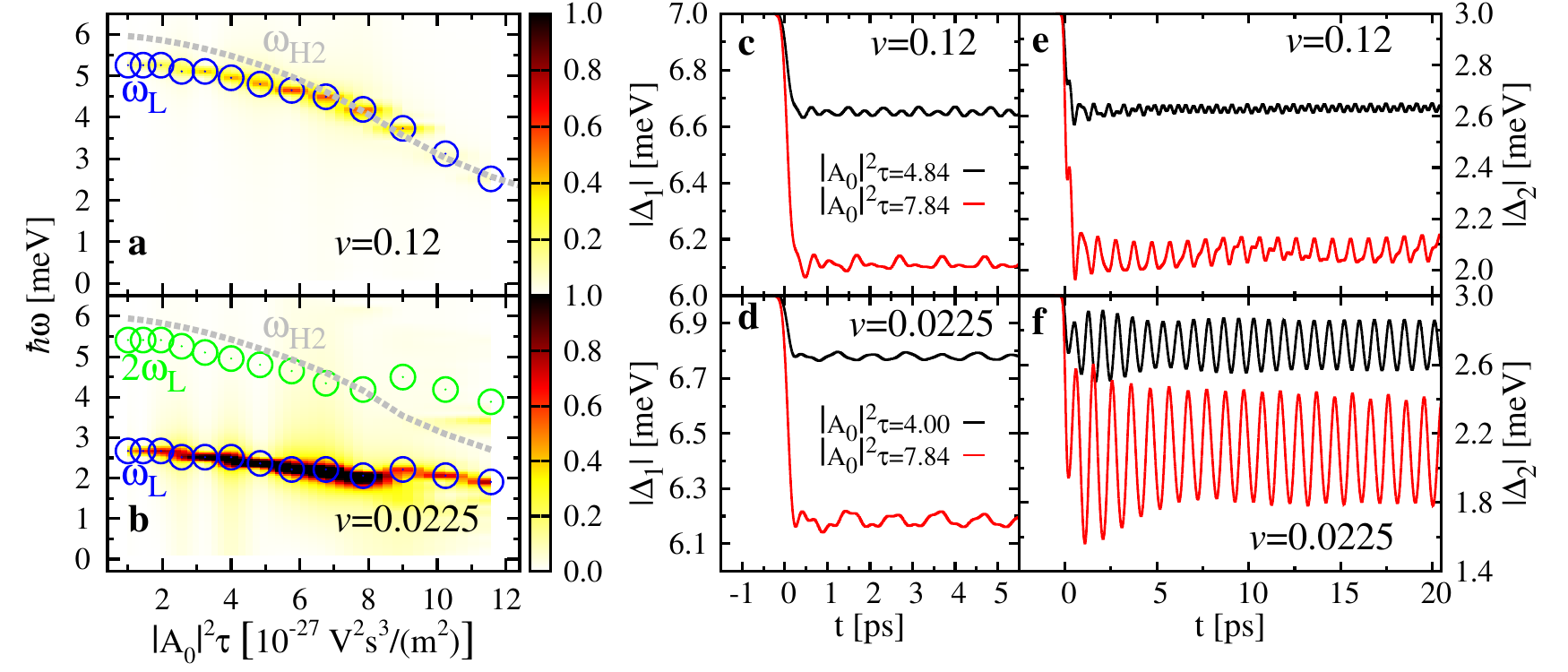}  
\caption{ \label{mFig5}
\textbf{Fluence dependence of gap dynamics.}
(\textbf{a}), (\textbf{b}) 
Fourier spectrum of the phase mode $\Phi_1 - \Phi_2$ as a function of  
laser fluence (integrated pulse intensity) $|{\bf A}_0|^2 \tau$ 
for two different interband couplings $v$
with pulse energy $\hbar \omega_{0}=8$~meV and pulse width $\tau = 0.4$~ps.
The amplitude of the phase fluctuations is represented by the color scale with dark red
and light yellow indicated high and low amplitudes, respectively. 
The open circles mark the frequencies of the nonequilibrium Leggett mode
$\omega_{\textrm{L}}$ and its higher harmonic $2 \omega_{\textrm{L}}$.
The grey dotted lines
display  the Higgs-mode $\omega_{\text{H}2}$, which coincides with the boundary to the continuum of Bogoliubov quasiparticle excitations.
(\textbf{c}) -- (\textbf{f}) 
Gap amplitudes $| \Delta_1|$ and $| \Delta_2|$ as a function of time $t$
for two different interband couplings $v$ and integrated pulse intensities $|{\bf A}_0|^2 \tau$ .
}
\end{figure*}

\noindent
\textbf{Pump response.}
Pumping the two-band superconductor with a short laser pulse of legnth $\tau  \ll \tau_{\Delta}$
 excites a nonthermal distribution of Bogoliubov quasiparticles above the gaps $\Delta_i$, which 
 in turn leads to a rapid, non-adiabatic change in the free-energy landscape $\mathcal{F}$. As a result,
the collective modes of the superconductor start to oscillate about the new minima of 
$\mathcal{F}$. This is clearly visible in Fig.~\ref{mFig1}, which 
shows the temporal evolution of the gap amplitudes $| \Delta_i | $ and of the 
phase difference $\Phi_1 - \Phi_2$ between the two gaps.
From the Fourier spectra in Figs.~\ref{mFig1}(d)-(f) we can see that 
three different modes (and their higher harmonics) are excited
at the frequencies $\omega_{\text{H}1}$, $\omega_{\text{H}2}$, and $\omega_{\textrm{L}}$.
The two modes at  $\omega_{\text{H}1}$ and $\omega_{\text{H}2}$  only exist
in the dynamics of $\Delta_1 (t)$ and  $\Delta_2 (t)$, respectively,
and their peaks are  
located at the energy of the superconducting gaps $\omega_{\text{H}i} = 2 | \Delta^{\infty}_i | / \hbar$, where $\Delta^{\infty}_i$ denotes the asymptotic gap value~\cite{volkov1974,amin2004,Schny11,Papen07,Papen08,Yuz05,Yuz06}.
This holds for all 
parameter regimes, even as the  laser fluence is increased far beyond the linear absorption region (see Fig.~\ref{mFig5}).
We therefore assign the peaks at $\omega_{\text{H}1}$ and $\omega_{\text{H}2}$ to the Higgs amplitude modes
of the two gaps. The higher Higgs mode $\omega_{\text{H}1}$ is strongly damped, because
it lies within the continuum of Bogoliubov quasiparticle excitations, which is lower bounded by $2 \Delta^{\infty}_2$.
For the lower mode $\omega_{\text{H}2}$, on the other hand,  the decay channel to quasiparticles is small, since
$\omega_{\text{H}2}$ is  at the continuum threshold. This is similar to the nonequilibrium Higgs mode of the single-gap superconductor NbN,
whose oscillations have recently been observed 
over a time period of about 10~ps by pump-probe measurements~\cite{matsunagaPRL12,matsunagaPRL13}.

%%%%%%%%%%%%%%%%%%%
%%%%%%%%%%%%%%%%%%%

Interestingly, two-band superconductors exhibit a third collective mode besides the two Higgs modes 
 at  a frequency $\omega_{\textrm{L}}$ below 
the quasiparticle continuum.
 This mode is most clearly visible in the dynamics of 
the phase difference $\Phi_1 - \Phi_2$ [Fig.~\ref{mFig1}(a)] and displays a striking dependence on interband coupling strength $v$.
With decreasing $v$ its frequency rapidly decreases, while its intensity grows. In 
the limit of  vanishing $v$, however, the third mode $\omega_{\textrm{L}}$ is completely absent.
We thus identify  $\omega_{\textrm{L}}$ as the Leggett phase mode, i.e., as
equal but opposite oscillatory phase shifts of the two coupled gaps.
Remarkably,   the Leggett phase mode is also observable in the time dependence
 of the gap amplitudes $\Delta_1(t)$ and $\Delta_2(t)$ [Figs.~\ref{mFig1}(b) and~\ref{mFig1}(c)], which indicates that  Higgs and Leggett modes 
 are coupled in  nonequilibrium superconductors.

To obtain a more detailed picture, we plot in Figs.~\ref{mFig3} and~\ref{mFig4} 
the energies of the amplitude and phase mode oscillations against the relative  interband coupling $v$.
This reveals that for small $v$ the nonequilibrium Leggett mode $\omega_{\textrm{L}}$ 
shows a square root increase,
which is in good agreement  
with the equilibrium Leggett frequency~\cite{Leggett66,sharapovEPJB}
\begin{eqnarray} \label{equilibriumLeggett}
\omega^0_{\textrm{L}}
=
2
\sqrt{
\Delta_1^{\infty} \Delta_2^{\infty}
\frac{ v  }
{V_{1} V_{2} - v^2 �}
\left( \frac{ 1}{\rho_1} + \frac{1}{\rho_2 } \right)
} ,
\end{eqnarray}
where $\rho_1$ and $\rho_2$ denote the density of states on the two bands.
Indeed, as displayed by the inset of Fig.~\ref{mFig3}, 
Eq.~\eqref{equilibriumLeggett} represents  an excellent parameter-free fit to the numerical data at low $v$.
For larger $v$, on the other hand, the nonequilibrium Leggett mode deviates from the square root behavior of 
Eq.~\eqref{equilibriumLeggett}.
That is,  as $\omega_{\textrm{L}}$ 
approaches the Bogoliubov quasiparticle continuum,  
it is repelled by the lower Higgs mode $\omega_{\text{H}2}$,  
evidencing a strong coupling between them.
As a result, the nonequilibrium Leggett mode is pushed below the
continuum and remains nearly  undamped for a wide range of $v$,
which is considerably broader than in equilibrium.
Moreover, due to the dynamical coupling among the collective modes,
$\omega_{\textrm{L}}$ and its higher harmonics
are observable not only in the phase difference $\Phi_1 - \Phi_2$,
but also in the dynamics of the gap amplitudes $\Delta_i (t)$ (blue and green circles in Fig.~\ref{mFig4}).

A key advantage of measuring collective modes by pump-probe experiments,
is that the frequencies of the Higgs modes can be adjusted by the pump fluence.
This is demonstrated in Fig.~\ref{mFig5}, which plots the dynamics of  
$\Delta_i (t)$  and $\Phi_1 - \Phi_2$ as a function of integrated pump pulse intensity $|{\bf A}_0|^2 \tau$.
With increasing pump fluence, more Cooper pairs are broken
up and superconductivity is more and more suppressed, as reflected in the reduction of the gap amplitudes.
At the same time, the frequency of the Higgs oscillations decreases,
since it is controlled by the superconducting gaps  after pumping.
Hence, 
it is possible to tune the lower  Higgs mode $\omega_{\text{H}2}$ to resonance with $\omega_{\textrm{L}}$,
 which strongly enhances  the magnitude of the collective-mode oscillations
[Figs.~\ref{mFig5}(a), \ref{mFig5}(c), and \ref{mFig5}(e)].
A similar enhancement is obtained when $\omega_{\text{H}2}$ is brought into resonance with twice the frequency of the Leggett mode
[Figs.~\ref{mFig5}(b), \ref{mFig5}(d), and \ref{mFig5}(f)].

\noindent
\textbf{Pump-probe signal.} 
Finally, let us discuss how the Higgs and Leggett modes can be observed in  pump-probe spectroscopy.
In view of the recent THz pump-THz probe experiments of Refs.~\cite{matsunagaPRL12,matsunagaPRL13,Matsunagascience14},
we focus on the dynamics of the optical pump-probe conductivity 
$  \sigma (\delta t, \omega) = j ( \delta t, \omega) / [ i \omega A ( \delta t, \omega)]$,
where $\delta t$ is the delay time between pump and probe pulses, $j ( \delta t, \omega)$ denotes the current density, and $A ( \delta t, \omega)$ 
represents the vector potential of the probe pulse. 
Since the probe pulse has a weak intensity, we neglect terms of 
second order and higher in the probe field $A ( \delta t, \omega)$.
Similar to recent experiments~\cite{matsunagaPRL12,matsunagaPRL13,Matsunagascience14,DalScience2013,mansartPNAS13,Pashkinscience14}, we take 
the probe pulse to be very short with width $\tau_{\textrm{pr}}=0.15$~ps
and center frequency  $\hbar\omega_{\textrm{pr}}$=5.5 meV (see Methods).
With this choice, the probe pulse has a broad spectral bandwidth 
such that the dynamics of the superconductor is probed over a very wide frequency range.

%%%%%%%%%%%%%%%%%%%
\begin{figure}[t!]
\includegraphics[width=0.49\textwidth]{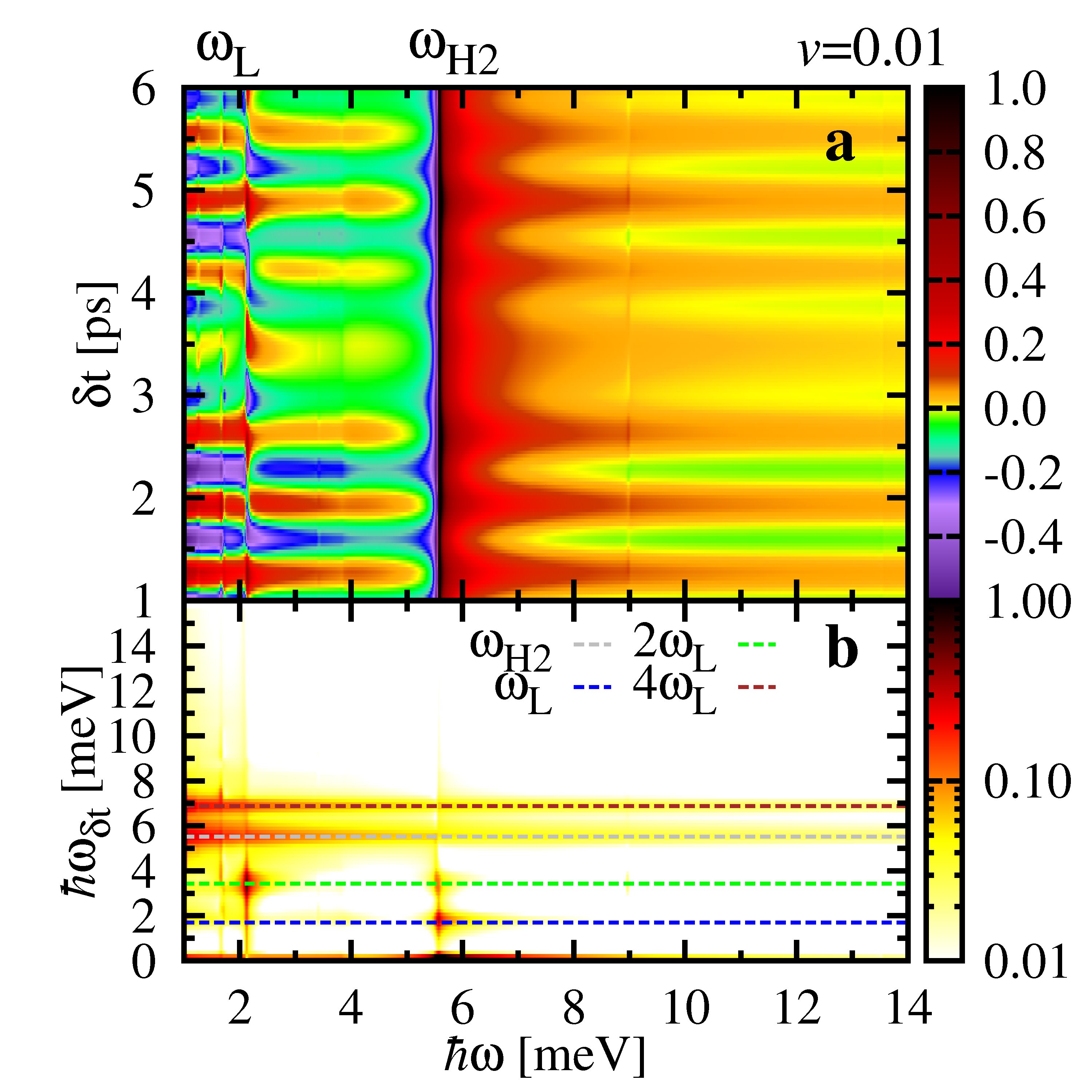} 
\caption{ \label{mFig6}
\textbf{Pump-probe spectrum.}
(\textbf{a}) Temporal evolution of the real part   
of the pump-probe response $\mathrm{Re}[ \sigma (\delta t, \omega)]$ for a two-band superconductor
excited by the same pump pulse as in Fig.~\ref{mFig1}.
The intensity of the pump-probe signal is represented by the color scale with dark red and dark violet indicating
high and low intensities, respectively.  Both the energy of the nonqeuilibrium Leggett mode $\omega_{\textrm{L}}$ and
the Higgs-mode $\omega_{\text{H}2}$ are visible in the pump-probe signal as sharp peaks.
(\textbf{b})~Fourier spectrum of the pump-probe signal. 
The pump-probe spectrum oscillates as 
a function of pump-probe delay time $\delta t$ with
the frequencies of the nonequilibrium Higgs mode  $\omega_{\text{H}2}$ (dashed gray),
 the frequency of the nonequilibirum Leggett mode $\omega_{\textrm{L}}$ (dashed blue), 
and the frequencies of higher harmonics of the Leggett mode
$2\omega_{\textrm{L}}$ (dashed green) and $4\omega_{\textrm{L}}$ (dashed red).
}
\end{figure}
%%%%%%%%%%%%%%%%%%%

In Fig.~\ref{mFig6}(a) we plot the real part of the pump-probe response  $\mathrm{Re}[\sigma ( \delta t, \omega)]$
versus delay time $\delta t$ and frequency $\omega$.
Clear oscillations are seen as a function of delay time $\delta t$.
These  are most prominent at the frequencies of the lower Higgs  
and the Leggett modes, $\omega_{\text{H}2}$ and $\omega_{\textrm{L}}$,
where $ \sigma ( \delta t, \omega)$ displays sharp edges as a function of $\omega$.
Fourier transforming with respect to $\delta t$ shows that
the dominant oscillations are $\omega_{\text{H}2}$ and $\omega_{\textrm{L}}$ (and its higher harmonics) [Fig.~\ref{mFig6}(b)].
We therefore predict that both the lower Higgs mode $\omega_{\text{H}2}$ and
the Leggett mode $\omega_{\textrm{L}}$ can be observed in THz pump-THz probe experiments
as oscillations of the pump-probe conductivity, in particular at the gap edge $2 \Delta^{\infty}_2 / \hbar$
and the Leggett mode frequency $\omega_{\textrm{L}}$.
The higher Higgs mode $\omega_{\text{H}1}$, on the other hand,
is not visible in the pump-probe signal, since it is strongly damped
by the two-particle continuum.

\noindent
\textbf{\large Discussion}\\
Using a semi-numerical method based on the density matrix approach, we have studied the 
non-equilibrium excitation of Higgs and Leggett modes in two-band superconductors.
While the amplitude Higgs and the Leggett phase mode are decoupled in equilibrium, we find that the out-of-equilibrium excitation process leads to a strong coupling between these two collective modes. As a result, the Leggett phase mode $\omega_{\textrm{L}}$ is pushed below the Bogoliubov quasiparticle continuum and remains undamped for a wide range of interband couplings (Figs.~\ref{mFig3} and \ref{mFig4}). Likewise, the lower Higgs mode $\omega_{\textrm{H2}}$ is only weakly damped, since its frequency is at the threshold to the quasiparticle continuum. 
In order to maximize the oscillatory signal of these collective modes in the pump-probe spectra,
it is necessary to choose the experimental parameters as follows:
(i) the pump-pulse duration $\tau$ should be smaller than the intrinsic response time of the 
superconductor $h / (2 | \Delta_i|)$, such that the collective modes are excited in a non-adiabatic fashion;
(ii) the pump-pulse energy needs to be of the order of the superconducting gap (i.e., in the terahertz regime),
so that Bogoliubov quasiparticles are excited across the gap, but modes at higher energies $\hbar \omega \gg  | \Delta_i|$ 
are not populated; and (iii) the pump-pulse intensity must not exceed a few nJ/cm$^2$ to ensure 
that the superconducting condensate is only partially broken up, but not completely destroyed. 
We have predicted that under these conditions both Higgs and Leggett modes
can be observed as clear oscillations in the time-resolved pump-probe absorption spectra (Fig.~\ref{mFig6}).
Similarly, we expect that collective mode oscillations  are  visible in
other pump-probe-type experiments, for example in time-resolved photoemission spectroscopy or time-resolved Raman scattering.

It would be intriguing to extend the present study to 
unconventional exotic superconductors, where several competing orders are present, such as
 heavy fermion superconductors or high-temperature cuprate and pnictide superconductors.
In these systems the pump pulse could be used to induce a transition from one competing order
to another. Furthermore, the unconventional pairing symmetries of these superconductors, such as 
the $d_{x^2-y^2}$-wave pairing of the cuprates, give rise to a multitude of new Higgs modes~\cite{barlasVarmaPRB_13}.
Our work  indicates that non-adiabatic excitation processes will induce interactions among these novel Higgs modes,
which await to be further explored both theoretically and experimentally.

\vspace{0.5cm}

\noindent
\textbf{\large Methods}\\
\noindent
\textbf{Model definition.} 
The gap equations for the BCS Hamiltonian $H_{\textrm{BCS}}$ (see Eq.~\eqref{eq_ham} in the main text)
are given by~\cite{Suhl59}
\begin{align}
\Delta_{1} &=\sum_{\textbf{k}'\in \mathcal{W}}\left(
V_{1} \langle c^{\phantom{\dagger}}_{-\textbf{k}',\downarrow,1}c^{\phantom{\dagger}}_{\textbf{k}',\uparrow,1} \rangle
+V_{12} \langle c^{\phantom{\dagger}}_{-\textbf{k}',\downarrow,2}c^{\phantom{\dagger}}_{\textbf{k}',\uparrow,2} \rangle\right) ,
\nonumber\\
\Delta_{2} &=\sum_{\textbf{k}'\in \mathcal{W}}\left(
V_{2} \langle c^{\phantom{\dagger}}_{-\textbf{k}',\downarrow,2}c^{\phantom{\dagger}}_{\textbf{k}',\uparrow,2} \rangle
+V_{12} \langle c^{\phantom{\dagger}}_{-\textbf{k}',\downarrow,1}c^{\phantom{\dagger}}_{\textbf{k}',\uparrow,1} \rangle\right) ,
\end{align}
where $V_1$ and $V_2$ denote the intraband interactions and $V_{12} = v V_1$ is the interband coupling.
The two-band superconductor is brought out of equilibrium via the coupling to a pump pulse, which is modeled by
\begin{align}
\label{eq:H_laser}
H_{\text{Laser}}
=
&\frac{e\hbar}{2}\sum_{\textbf{k},\textbf{q},\sigma,l}\frac{(2\textbf{k}+\textbf{q})\textbf{A}_{\textbf{q}}(t)}{m_l}c^{\dagger}_{\textbf{k}+\textbf{q},\sigma,l}c^{\phantom{\dagger}}_{\textbf{k},\sigma,l}\\
&+\frac{e^2}{2}\sum_{\textbf{k},\textbf{q},\sigma,l}\frac{\left(\sum_{\textbf{q}^{\prime}}\textbf{A}_{\textbf{q}-\textbf{q}^{\prime}}(t)\textbf{A}_{\textbf{q}^{\prime}}(t)\right)}{m_l}c^{\dagger}_{\textbf{k}+\textbf{q},\sigma,l}c^{\phantom{\dagger}}_{\textbf{k},\sigma,l}\notag,
\end{align}
where $m_l$ is the effective electron mass of the $l$th band and $\textbf{A}_{\textbf{q}}(t)$ represents the transverse vector potential of the pump laser. We consider a Gaussian pump pulse described by
\begin{align}
\textbf{A}_{\textbf{q}}(t)=\textbf{A}_{0}  e^{-\left(\frac{2\sqrt{\ln{2}}t}{\tau }\right)^2}\left(\delta_{\textbf{q},\textbf{q}_{0} }e^{-i\omega_{0}  t}+\delta_{\textbf{q},-\textbf{q}_{0} }e^{i\omega_{0}  t}\right), 
\end{align}
with central frequency $\omega_{0} $, pulse width $\tau$,
 light-field amplitude $\textbf{A}_{0}= | \textbf{A}_{0} | \hat{\bf e}_y$,
and photon wave-vector ${\bf q}_0 = q_0 \hat{\bf e}_x$.\\

\vspace{0.2cm}

\noindent
\textbf{Density matrix formalism.} 
In order to simulate the nonequilibrium dynamics of the two-band superconductor~\eqref{eq_ham}, we use a semi-numerical method
based on the density matrix formalism. This approach involves the analytical derivation 
of equations of motions for the Bogoliubov quasiparticle densities
$\langle\alpha^{\dagger}_{\textbf{k},l}\alpha^{\phantom{\dagger}}_{\textbf{k}',l } \rangle$, 
$\langle\beta^{\dagger}_{\textbf{k},l}\beta^{\phantom{\dagger}}_{\textbf{k}',l } \rangle$,
$\langle\alpha^{\dagger}_{\textbf{k},l}\beta^{\dagger}_{\textbf{k}',l } \rangle$,
 and $\langle\alpha^{\phantom{\dagger}}_{\textbf{k},l}\beta^{\phantom{\dagger}}_{\textbf{k}',l } \rangle$,
which are then integrated up numerically using a Runge Kutta algorithm. 
The Bogoliubov quasiparticle densities are defined in terms of
the fermionic operators $\alpha_{\textbf{k},l}$ and $\beta_{\textbf{k},l}$, with
\begin{align}
\alpha_{\textbf{k},l}=u_{\textbf{k},l}c^{\phantom{\dagger}}_{\textbf{k},l,\uparrow}-v_{\textbf{k},l}c^{\dagger}_{\textbf{-k},l,\downarrow}, 
\\
\beta_{\textbf{k},l} =v_{\textbf{k},l} c^{\dagger}_{\textbf{k},l,\uparrow}+ u_{\textbf{k},l} c^{\phantom{\dagger}}_{-\textbf{k},l,\downarrow} ,
\end{align}
where 
$
v_{\textbf{k},l}= \Delta_l (t_i ) / |\Delta_l (t_i ) | \sqrt{ ( 1 - \epsilon_{\textbf{k},l} /  E_{\textbf{k},l}   )/2 }
$,
$
u_{\textbf{k},l}=\sqrt{  ( 1+ \epsilon_{\textbf{k},l} /  E_{\textbf{k},l}  )/2}
$,
and $E_{\textbf{k},l}=\sqrt{\epsilon^2_{\textbf{k},l}+|\Delta_l (t_i ) |^2}$.
We emphasize that the coefficients $u_{\textbf{k},l}$ and $v_{\textbf{k},l}$ do not depend on time, i.e.,
the temporal evolution of the quasiparticle densities is computed with respect to a fixed time-independent
Bogoliubov-de Gennes basis in which the initial state is diagonal.
The equations of motion for the quasiparticle densities are readily obtained from Heisenberg's equation of motion.
Since Eq.~\eqref{eq_ham} represents a mean-field Hamiltonian, this yields a closed system of differential
equations, and hence no  truncation is needed (for details see, e.g., Refs.~\cite{Papen07,Akba13,Krull14,Krullphd}).

\vspace{0.2cm}

\noindent
\textbf{Pump-probe response.} 
All physical observables, such as the current density  ${\bf j}_{{\bf q}_{\textrm{pr}}} (\delta t, t)$, can be expressed in terms of the  
 quasiparticle densities.
For the current density we find that
\begin{eqnarray}
&&
{\bf j}_{\textbf{q}_{\text{pr}}}(\delta t, t)
=
\frac{-e\hbar}{2mV}
\sum\limits_{\textbf{k},l, \sigma}(2\textbf{k}+\textbf{q}_{\text{pr}})
\left\langle c^{\dagger}_{\textbf{k},l,\sigma}c^{\phantom{\dagger}}_{\textbf{k}+\textbf{q}_{\text{pr}},l,\sigma}\right\rangle (\delta t, t) 
\nonumber\\
&&
\; \qquad - \frac{e^2}{m V}
\sum\limits_{\textbf{k},l, \textbf{q}, \sigma}
{\bf A}_{\textbf{q}_{\text{pr}} - \textbf{q}} \left\langle c^{\dagger}_{\textbf{k},l,\sigma}c^{\phantom{\dagger}}_{\textbf{k}+\textbf{q},l ,\sigma}\right\rangle (\delta t, t)  ,  \;
\end{eqnarray}
where 
${\bf A}_{{\bf q}_{\textrm{pr}}} (\delta t, t)$
and 
${\bf q}_{\textrm{pr}}= \left| {\bf q}_{\textrm{pr}} \right| \hat{{\bf e}}_x$
are the vector potential and
 the wave vector of the probe pulse, respectively.
With this, we  obtain the pump-probe conductivity via~\cite{Krull14,2015arXiv150701200S}
\begin{align}
\sigma(\delta t , \omega)=\frac{j(\delta t, \omega)}{i\omega A( \delta t , \omega)},
\end{align}
where 
$
j ( \delta t,  \omega )
$
and 
$
A ( \delta t, \omega ) 
$
denote the Fourier transformed $y$~components of the current density ${\bf j}_{{\bf q}_{\textrm{pr}}}(\delta t, t)$ and the vector potential
${\bf A}_{{\bf q}_{\textrm{pr}}} (\delta t, t)$, respectively. 
To compute the effects of the probe pulse, we neglect 
terms of second order and higher in the probe field $A_{\textrm{pr}}(t)$,
since the probe pulse has a very weak intensity.

\vspace{0.2cm}

\noindent
\textbf{Numerical discretization and integration.} 
 To keep the number of equations of motion manageable, we
have to restrict the number of considered points in momentum space. The first restriction is that we only take expectation values with indices $\textbf{k}$ and $\textbf{k}+\textbf{q}\in \mathcal{W}$ into account. This means that we concentrate on the $\textbf{k}$-values where the attractive pairing interaction takes place.
Furthermore, since the external electromagnetic field may  add or subtract only momentum $n \textbf{q}_0$,
it is sufficient to consider   expectation values with indices ($\textbf{k}$, $\textbf{k}+n\textbf{q}_0$), where $n \in \mathbb{Z}$.
For small amplitudes $|\textbf{A}_{\textrm{q}_0}|$ the off-diagonal elements of the quasiparticle densities decrease rapidly as $n$ increases, since $(\textbf{k}$, $\textbf{k}+n\textbf{q}_0)=O(|\textbf{A}_{\textrm{q}_0}|^{|n|})$. 
Thus, we set all entries with $n>4$ to zero. 
With this momentum-space discretization, we obtain of the order
of  $10^{5}$ equations, which we 
 are able to solve numerically using high-efficiency parallelization. 
Further, technical details can be found in Refs.~\cite{Krull14,Krullphd}.

\vspace{0.5 cm}
 
\noindent
\textbf{Acknowledgements} \\
 We gratefully acknowledge many useful discussions with A. Avella, S. Kaiser and R. Shimano. G.S.U. and H.K. acknowledge financial support by the DFG in TRR 160.
 H.K.\ thanks the Max-Planck-Institut FKF Stuttgart for its hospitality.

\vspace{0.5 cm}
 
\noindent
\textbf{Competing financial interests} \\
 The authors declare that they have no
competing financial interests.

\vspace{0.5 cm}
 
\noindent
\textbf{Author contributions} \\
 The density matrix simulations were developed and run by H.K., N.B., and A.P.S. 
 All authors contributed to the discussion and interpretation of the results and to the writing of the paper.

 %%%%%%%%%%%%%%%%%%%%%%%%%%%%%%%%%%
 %%%%%%%%%%%%%%%%%%%%%%%%%%%%%%%%%%

\end{document}